\documentclass[12pt]{article}
\usepackage{graphicx}
\usepackage{epsfig}
\usepackage[figuresright]{rotating}
\usepackage{amsmath}
\usepackage{amssymb}
\usepackage{amsfonts}
\usepackage{array}

\newcommand{\pibf}{\mbox{\boldmath $\pi$}}
\newcommand{\taubf}{\mbox{\boldmath $\tau$}}

\newcommand{\vectau}{\mbox{\boldmath$\tau$}}

\renewcommand{\thefootnote}{\fnsymbol{footnote}}
\newcommand{\fpartial}{\mbox{$/\!\!\!\partial$}}

\usepackage{cite}
\usepackage{amssymb}
\usepackage{amsfonts}

\begin{document}

\begin{center}
{\LARGE{\bf Nambu's Nobel Prize, the $\sigma$ meson 
and the mass  of visible matter}}\\
[1ex] 
Martin Schumacher\footnote{mschuma3@gwdg.de}\\
II.  Physikalisches Institut der Universit\"at G\"ottingen,
Friedrich-Hund-Platz 1\\ D-37077 G\"ottingen, Germany
\end{center}

\begin{abstract}
The electroweak Higgs boson has been discovered in
ongoing experiments at the LHC, leading to a mass of this particle of
126 GeV.  This Higgs boson mediates the generation of mass for
elementary particles, including the mass of
elementary (current) quarks. These current-quark masses leave 98\% of the 
mass of the atom unexplained. This large fraction is  mediated by strong 
interaction,
where instead of the Higgs boson the $\sigma$ meson is the mediating particle.
Though already introduced in 1957 by Schwinger,
the $\sigma$ meson has been integrated out in many theories of hadron properties
because it had not been observed and was doubted to exist. With the 
observation of the $\sigma$ meson in recent experiments on Compton scattering 
by the nucleon at MAMI (Mainz) it has become timely to review the   status
of experimental and theoretical researches  on this topic.
\end{abstract}

\renewcommand{\thefootnote}{\arabic{footnote}}

\section{Introduction}

Modern cosmology teaches us that matter is composed of three contributions,\\
(i) dark energy, amounting to 68.3\% of the total matter (see e.g. 
\cite{PDG}),\\
(ii) dark matter, amounting to 26.8\% of the total matter 
(see e.g \cite{PDG}).and\\
(iii) normal or visible matter amounting to 4.9\% of the total matter.\\
Dark energy is related to  space and has the property of expanding
the universe with increasing velocity. Dark matter is observed
in the dynamics of galaxies. The spiral arms of galaxies at large distances
from the center move with much larger velocities  than expected from 
Kepler's third law  if only normal matter is taken into account. With 
additional
matter which only interacts via gravitational  forces  Kepler's third
law can be 
preserved. This leads to the notion of dark matter which also shows up in
other phenomena, e.g. gravitational lenses.

Normal or visible matter consists of atoms, ions and electrons   where the by
far  
largest part of the mass is provided by  nucleons in free space or bound
in  nuclei. The 
nucleons consist of three quarks having current-quark masses of 
$m^0_{\rm u} \sim 3 - 5$ MeV and  $m^0_{\rm d} \sim 7 - 9$ MeV for the up 
quark and the down quark,
respectively, where the ranges take into account that these masses are scale
dependent. These current-quark masses are the ones which enter into the
standard model and, thus, are attributed to the Higgs boson. This means that
the Higgs boson is responsible for only 2\% of the visible matter, whereas
98\% are missing. There are different explanations for the missing 98\% 
of the mass of the nucleon. One frequently cited explanation makes use of 
the total  energies of the current quarks and of the gluon content 
of the nucleon  where the nucleon mass is generated 
via computer simulation on the lattice (see e.g. \cite{alvarez13}).  
These explanations disregard
that there are much more appropriate explanations, which make use of effective
theories describing the low-energy limit of QCD. These effective theories make
use of the
$\sigma$ meson, a low-mass sibling of the Higgs boson, which creates 
masses   of the constituent quarks of 326 MeV in the chiral limit, i.e.
for the case that the effects of the Higgs field are turned off. Adding these
masses together a nucleon mass of $\sim 980$ MeV is obtained which
is increased to $\sim 1$ GeV after  the Higgs field is turned on.
This predicted nucleon mass is larger than the observed nucleon mass 
by $\sim 20$ MeV per constituent quark. The necessary subtraction 
of $\sim 20$ MeV per constituent quark corresponds
to the binding energy of 
 $\sim 8$  MeV per nucleon observed in the nucleus on the average.
This tentative interpretation is reasonable because the mesonic component
of the forces between constituent quarks in a nucleon and nuclear forces
are of the same origin \cite{glozman96}. In the nucleon an additional 
component due to gluon exchange is expected.

The 2013 Nobel  Prize in   Physics has been awarded to Fran\c{c}ois Englert
and Peter Higgs "for the theoretical discovery of a mechanism that contributes 
to our understanding of the origin of mass of subatomic particles, and which 
recently was confirmed through the discovery of the predicted fundamental
 particle, by the ATLAS and CMS experiments at CERN's Large Hadron Collider".
This award followed a related one where
one half of the 2008  Nobel Prize in Physics  was awarded to Yoichiro Nambu
``for the discovery of the mechanism of spontaneous broken symmetry in
subatomic physics''. In order to understand the importance of  this latter
award it is very important to read the Nobel Lecture \cite{nambu09}
which was presented by Giovanni Jona-Lasinio, a younger coauthor
of the   famous Nambu--Jona-Lasinio model \cite{nambu61}
published in 1961. This Nobel
Lecture having the title ``Spontaneous symmetry breaking in particle physics:
A case of cross fertilization'' describes the way from  superconductivity
to particle physics which led to the  Nambu--Jona-Lasinio (NJL)
Lagrangian
\begin{equation}
{\cal L}=-\bar{\psi} \gamma^{\mu}  \partial_\mu \psi +g [ (\bar{\psi}\psi)^2
-\Sigma_i(\bar{\psi}\gamma_5 \taubf_i \psi)(\bar{\psi} \gamma_5 \taubf_i
\psi) ].       
\end{equation}
As pointed out in the Nobel Lecture
this equation describes an isovector $0^-$ pion and an isoscalar $0^+$.
In the standard model of particle physics, the NJL model may be
regarded as an effective theory for the QCD with respect to generation of the
so-called constituent masses. In analogy to different descriptions of
superconductivity the NJL model goes over to the linear $\sigma$ model
(L$\sigma$M) of Gell-Mann and L\'evy \cite{gellmann60}. In the Nobel Lecture
Nambu makes the important statement: "If this analogy turns
out real, the Higgs field  might be an effective description of the underlying
dynamics."  
 This Higgs field  is the Higgs field of strong interaction represented by the 
$\sigma$ meson.

An explanation of the constituent-quark mass in terms of symmetry breaking
mediated by the $\sigma$ meson remained uncertain as long as the $\sigma$
meson had not been observed.
This, however, has changed dramatically  in the last  years after 
the $\sigma$ meson has been observed as part of the  constituent-quark 
structure via Compton scattering by the nucleon. This experiment was carried
out at MAMI (Mainz) and published in 2001 \cite{galler01,wolf01}. 
The  final interpretation of the 
results obtained required some further theoretical studies which were
published in 2010 \cite{schumacher10} , 2011 \cite{schumacher11}  and 
2013 \cite{schumacher13}.
Through this experimental and theoretical work the 
$\sigma$ meson is by now well investigated and the process of mass 
generation of constituent quarks well understood. Additional aspects of the
structure of the $\sigma$ meson following from $\gamma\gamma\to\pi\pi$
and $\pi\pi\to\pi\pi$ reactions are discussed in section 5.4 of Ref.
\cite{schumacher13}.

For the Higgs boson the theoretical  research 
started with the work of Goldstone (1961 \cite{goldstone61} 
and 1962 \cite{goldstone62} ) where it was shown
that spontaneous symmetry breaking leads to massless particles in addition to
a heavy scalar particle. This is not a problem for the $\sigma$ meson where the
light $\pi$ mesons are massless in the chiral limit and have only a small
mass as real particles,  serving as pseudo-Goldstone bosons. For the 
Higgs boson these massless Goldstone bosons are strongly unwanted particles 
because they seem not to be present in nature. Therefore, 
in a number of papers 
scenarios were developed leading to symmetry breaking without Goldstone
bosons. This essential modification is related to the introduction
of massless gauge boson which swallow the Goldsone bosons and in this
way generate mass and a longitudinal field component. 
In parallel to the work of Higgs \cite{higgs64a,higgs64b}
two other groups, Englert-Brout 
\cite{englert64} and Guralnik-Hagen-Kibble \cite{guralnik64} worked on
electroweak symmetry breaking, so that -- in the order of priority -- the
mechanism of 
mass generation of the electroweak gauge boson was temporarily called the 
Englert-Brout---Higgs---Guralnik-Hagen-Kibble mechanism. 
There are ongoing experiments at the LHC  where the mass of the supposed
 Higgs boson was determined in 2012 to be 126 GeV. The related Nobel Prize
was awarded in 2013 to only two researchers, Englert and Higgs, because of
priority and because Brout passed away shortly before. The now official
name of the mechanism is Brout-Englert-Higgs (BEH) mechanism.

Nowadays the origin of the theory of spontaneous symmetry breaking is most 
frequently 
attributed to  the work of Peter Higgs \cite{higgs64a,higgs64b,higgs66}.
But it was  Peter Higgs himself who correctly pointed out  \cite{higgs66}, 
that vacuum expectation 
values of scalar fields  might play a role in breaking of 
symmetries
was first noted by Schwinger \cite{schwinger57}.
This means that strong and electroweak symmetry
breaking both can be traced back to the seminal work of Schwinger
\cite{schwinger57} and that the introduction of the $\sigma$ meson
inspired the electroweak symmetry breaking, though these two processes take
place at  completely different scales. The interest in this 
interplay between the two sectors of symmetry breaking is of importance up to
the present.

In the following section Schwinger's seminal ideas are described.
Thereafter it is shown how the NJL model is related to the
L$\sigma$M. Furthermore, the formal equivalence of the L$\sigma$M and the
electroweak Glashow-Weinberg-Salam (GWS) theory  of the standard model  
is briefly pointed out.
This formal equivalence is complete only for the scalar Higgs ($\sigma$)
part but not for the Goldstone boson part.  As stated above, in the
electroweak GWS
theory the Goldstone bosons do not exist as observable particles but
are transferred into the longitudinal components
of the electroweak gauge bosons via the 
BEH mechanism and in this way give the electroweak gauge bosons a mass.  
In case of the NJL and the L$\sigma$M models the Goldstone bosons are
identical with the triplet of $\pi$ mesons and thus are explicitly existing
particles. 

\section{Schwinger's $\sigma$ meson}

In 1957 Schwinger \cite{schwinger57} was interested in 
the question why  particles of different masses exist
though from the point of view of theoretical
physics a World with massless particles would be much simpler or, 
in other words,  much more symmetric.   Schwinger was aware of the 
fact that this question was of fundamental
importance and, therefore, he gave his paper the title "A Theory of 
the Fundamental Interactions".  As an abstract he cites A. Einstein 
with the words "The axiomatic basis of theoretical physics cannot be 
extracted from experiments but must be freely invented". The purpose 
of the paper was  to present a description of the stock of elementary 
particles  within the framework of the theory of quantized fields,
which had been a subject of Schwinger's research in the years before
\cite{schwinger51}. 
In \cite{schwinger57} Schwinger  reconsiders the $\sigma$ meson accompanying 
the triplet of $\pi$ mesons, a topic  which he had first discussed 
in \cite{schwinger56}. 
The important point he is making is that 
the absence of such a stable or metastable particle does not necessarily 
mean that  a field of the corresponding type does not exist.
If the  $\sigma$ field were scalar, as contrasted with the known
pseudoscalar character of the $\pi$ field, and the $\sigma$ meson 
had a mass greater than 2$m_\pi$, then the  $\sigma$ meson 
would be highly unstable against rapid disintegration into two $\pi$ mesons.
Therefore, the $\sigma$ meson has a high probability to escape
experimental observation.

As a first  application the mass of the $\mu$ lepton is discussed by Schwinger \cite{schwinger57}. 
Within a theory of quantized fields the $\sigma$ meson corresponds to a 
$\sigma$ field $\phi_{(0)}$.
As a field which is scalar under all operations 
in the three-dimensional isotopic space and in space-time, $\phi_{(0)}$ 
has a nonvanishing expectation value $\langle \phi_{(0)}\rangle$. Then
one could at least anticipate that 
a suitable $\mu$-lepton
mass constant might emerge from
$g_\mu \langle \phi_{(0)}\rangle$ without requiring a particularly 
large coupling constant $g_\mu$. 

Translated into present-day language Schwinger introduced a generic equation
between the mass $m$ of a particle and the vacuum expectation value of a 
scalar field $\phi_{(0)}$ given in the form
\begin{equation}
m \propto g \langle \phi_{(0)} \rangle.
\label{generic}
\end{equation}
For the purpose of the present paper we translate Eq. (\ref{generic}) into
three 
related equations, {\it viz.}
\begin{equation}
m_l = g_{Hll}\frac{1}{\sqrt{2}} v,
\label{generic-1}
\end{equation}
\begin{equation}
m^0_q = g_{Hqq}\frac{1}{\sqrt{2}} v
\label{generic-2}
\end{equation}
and
\begin{equation}
m^{\rm cl}_q = g_{\sigma qq} f^{\rm cl}_\pi.
\label{generic-3}
\end{equation}
Eqs. (\ref{generic-1}) and  (\ref{generic-2})  relate the lepton 
and  current-quark masses, respectively,
to the
electroweak vacuum expectation value $v$  of the Higgs field and Eq. 
(\ref{generic-3}) the constituent quark mass in the chiral limit (cl), where
the 
effects of the Higgs field are turned off, to the pion decay constant
$f^{\rm cl}_\pi$ in the chiral limit. In case of
Eqs. (\ref{generic-1}) and (\ref{generic-2})  the  Higgs-lepton and
Higgs-quark coupling constants $g_{Hll}$ and 
$g_{Hqq}$ can be calculated from  the known electroweak vacuum expectation
value $v= 246$ GeV  and the known lepton mass $m_l$ and current-quark 
mass  $m^0_q$, respectively.  In case of
Eq. (\ref{generic-3}) the pion decay constant in the chiral limit is
$f^{\rm cl}_\pi= 89.8$ MeV and $g_{\sigma qq}=2\pi/\sqrt{3}=3.62$. More
details will be given later.

In a second application of the $\sigma$ meson Schwinger \cite{schwinger57}
discussed the
mass difference between the charged gauge bosons of weak interaction and the
photon. In this work Schwinger had almost all the ingredients to construct
the SU(2)$\times$U(1) electroweak theory. Schwinger suggested the problem 
to Sheldom Glashow for further investigation. This led to the work of Sheldon
Glashow  \cite{glashow61} six years before that of Steven Weinberg 
\cite{weinberg67}. Glashow, Weinberg and Abdus Salam \cite{salam69}  
 shared the Nobel Prize (1979) for the unification of weak interaction with
 electromagnetism. In this theory the masses of the three vector bosons
of weak interaction are generated via the  BEH mechanism, 
where the
Goldstone bosons of electroweak symmetry breaking are transformed into 
the longitudinal field components and thus giving these three vector
bosons a mass.

In (\ref{generic-1})   -- (\ref{generic-3}) two sources of mass are discussed,
{\it viz}  symmetry breaking mediated by the Higgs field 
(Eqs. \ref{generic-1} and \ref{generic-2})
and spontaneous symmetry breaking mediated by the $\sigma$ field (Eq. 
\ref{generic-3}). These are the
main sources  of mass generation. In case of strong interaction there are 
in addition dynamic effects related
to excited states,  the interaction of spins and effects due to gluons where
the latter effects show up in the form of glueballs and    the $U_A(1)$ 
anomaly.

\section{Summary on the  Nambu--Jona-Lasinio
  model and the linear $\sigma$ model}

The Lagrangians of the NJL model 
\cite{nambu61,lurie64,eguchi76,vogl91,klevansky92,hatsuda94,bijnens96}
for the number of flavors $N_f=2$ are given by
\begin{eqnarray}
&&{\cal L}_{\rm NJL}=\bar{\psi}(i\fpartial-m_0) \psi
+
\frac{G}{2}[(\bar{\psi}\psi)^2+(\bar{\psi}i\gamma_5\vectau\psi)^2],\label{ql2}
\\
&&{\cal L'}_ {\rm
  NJL}=\bar{\psi}i\fpartial\psi-g\bar{\psi}(\sigma+i\gamma_ 5
\vectau\cdot\pibf)\psi-\frac12\delta\mu^2(\sigma^2+\pibf^2)+\frac{gm_0}{G}
\sigma, \label{ql3}\\
&& G=g^2/\delta\mu^2, \quad 
\delta\mu^2=(m^{\rm cl}_\sigma)^2.
 \label{ql4a}
\end{eqnarray}
In Eq. (\ref{ql2}) the interaction between
the fermions is parameterized by the four-fermion interaction constant $G$
and explicit symmetry breaking by the average current-quark mass
$m_0=\frac12(m^0_u+m^0_d)$. 
Eq. (\ref{ql3}) differs from  (\ref{ql2}) by the fact that the interaction
is described by the exchange of bosons. This leads to the occurrence of a mass
counter term parameterized by $\delta\mu^2$. This mass counter term may 
be identified with the mass squared  $(m^{\rm cl}_\sigma)^2$
 of the $\sigma$ meson in the 
chiral limit.   Eq. (\ref{ql4a}) shows how
the two coupling constants $G$ and $g$ are related to each other.

The use of two versions of the NJL model has the advantage that for 
many applications the four-fermion version is more convenient whereas
the bosonized version describes the interaction of the constituent quark
with the QCD vacuum  through the exchange of the $(\pibf,\sigma)$ meson quartet
which we consider as the true description of the physical process.
In the chiral limit we may write 
\begin{equation}
G \rightarrow \frac{g^2}{(m^{\rm cl}_\sigma)^2-q^2}, \quad q^2 \rightarrow 0
\label{propagator}
\end{equation}
where the four-fermion coupling constant $G$ is replaced by a $\sigma$-meson
propagator in the limit $q^2 \rightarrow 0$, with $q$ being the momentum
carried by the $\sigma$ meson.

The bosonized version  may be obtained with the following ansatz \cite{vogl91}
\begin{eqnarray}
&&\sigma=-\frac{G}{g}\bar{\psi}\psi, \label{bos1}\\
&&\pibf=-\frac{G}{g}\bar{\psi}\,i\,\gamma_5\taubf\,\psi \label{bos2}
\end{eqnarray}
where the $\bar{q}q$ structure of the $(\pibf,\sigma)$ meson quartet is given
  by 
\begin{equation}
|\pi^+\rangle=u\bar{d},\quad|\pi^0\rangle=\frac{1}{\sqrt{2}}
(-u\bar{u}+ d\bar{d}  ),\quad|\pi^-\rangle=-d\bar{u},\quad |\sigma
\rangle=\frac{1}{\sqrt{2}}(u\bar{u}+d\bar{d}).
\label{qqstructure}
\end{equation}
The relation of the $q\bar{q}$ structure of the $\sigma$ meson
with other structures has been
investigated in \cite{schumacher11}.

The NJL model is not complete at this point and, accordingly,  has to be
supplemented. 
Using diagrammatic techniques the following  equations may be found 
\cite{klevansky92,hatsuda94}  for the non-strange $(\pibf,\sigma)$ sector
\begin{eqnarray}
&&M^*=m_0+ 8\, i\, G N_c \int^{\Lambda}\frac{d^4 p}{(2\pi)^4}
\frac{M^*}{p^2-M^{*2}},\quad M=-\frac{8\,iN_cg^2}{(m^{\rm cl}_\sigma)^2}
\int\frac{d^4p}{(2\pi)^4}\frac{M}{p^2-M^2},
\label{gapdiagram}\\
&&f^2_\pi = -4\,i\,N_cM^{*2} \int^{\Lambda}\frac{d^4p}{
(2\pi)^4}\frac{1}{(p^2-M^{*2})^2},\quad f^{\rm cl}_\pi=-4iN_cgM\int\frac{d^4p}{(2\pi)^4}
\frac{1}{(p^2-M^2)^2}.
\label{fpiexpress}
\end{eqnarray}
The expression given on the l.h.s. of (\ref{gapdiagram}) is the 
gap equation with $M^*$
being the mass of the constituent quark with the contribution 
$m_0$ of the
current quarks included. The r.h.s.
shows the gap equation 
for the nonstrange $(n\bar{n})$ constituent-quark mass $M$ in the chiral limit.
The l.h.s. of Eq.
(\ref{fpiexpress}) represents  the pion decay constant and the r.h.s.
the same quantity in the chiral limit. 
 For further details we refer to 
\cite{schumacher06,delbourgo95}.

Making use of dimensional regularization the
Delbourgo-Scadron \cite{delbourgo95} relation 
\begin{equation}
M=\frac{2\pi}{\sqrt{N_c}}f^{\rm cl}_\pi, \quad N_c=3
\label{sigmamass}
\end{equation}
may be obtained from 
the r.h.s of Eqs. (\ref{gapdiagram}) and (\ref{fpiexpress}).
This important relation shows that the mass of the constituent quark 
in the chiral limit and the pion decay constant in the chiral limit are
proportional to each other. This relation is valid independent of the
flavor content of the constituent quark, e.g. also for a constituent quark
where the  $d$-quark is replaced by a $s$-quark. Furthermore, it has been
shown \cite{delbourgo98,delbourgo02,scadron13} that (\ref{sigmamass}) is  
valid independent of the
regularization scheme.

From (\ref{sigmamass}) we obtain for the constituent-quark mass in the chiral
limit 
\begin{equation}
M=326 \,\,\, \text{MeV},\quad m^{\rm cl}_\sigma= 2 M = 652 \,\,\, \text{MeV}
\label{sigmamass-2}
\end{equation}
where use is made of $f^{\rm cl}_\pi=89.8$ MeV. The mass $m_\sigma$ of the
$\sigma$ meson including the effects of explicit symmetry breaking may be
obtained from  
\begin{equation}
m^2_\sigma= (m^{\rm cl}_\sigma)^2 + \hat{m}^2_\pi
\label{sigmamass-3}
\end{equation}
where $\hat{m}_\pi$ is the average pion mass.This equation has been derived in the NJL as well as in the L$\sigma$M. Numerically this leads to
\begin{equation}
m_\sigma= 666\,\,\, \text{MeV}.
\label{sigmamass-4}
\end{equation}

As discussed in Nambu's Nobel Lecture
it is of interest to compare the NJL model with the linear $\sigma$ model
(L$\sigma$M) \cite{gellmann60,alfaro73,donoghue96}.
The L$\sigma$M may be written in terms of three contributions, a fermion,
a boson and an explicit contribution:
\begin{eqnarray}
&&{\cal L}_{{\rm L}\sigma{\rm M}}^{\rm fermion}=
\bar{\psi}i\fpartial\psi- g\bar{\psi}(\sigma+i\gamma_ 5
\vectau\cdot\pibf)\psi,\label{fermion}\\
&&\quad\quad\quad\,\,=
\bar{\psi}i\fpartial\psi- g\bar{\psi}(\sigma'+i\gamma_ 5
\vectau\cdot\pibf)\psi+ gf^{\rm cl}_\pi \bar{\psi}\psi,
\label{fermionprime}\\
&&{\cal L}_{{\rm L}\sigma{\rm M}}^{\rm boson}=
\frac12\partial_\mu\pibf\cdot\partial^\mu
\pibf + \frac12 \partial_\mu\sigma\partial^\mu\sigma 
+\frac{\mu^2}{2}(
\sigma^2+\pibf^2)-\frac{\lambda}{4}(\sigma^2+\pibf^2)^2, \label{boson}\\
&&{\cal L}_{{\rm L}\sigma{\rm M}}^{\rm explicit}=
f^{\rm cl}_\pi m^2_\pi 
\sigma. \label{ql1}
\end{eqnarray}
Eq. (\ref{fermion}) is the fermion part of the L$\sigma$M Lagrangian
formulated on the quark level, Eq. (\ref{fermionprime}) the same relation
after the shift of the $\sigma$ field
\begin{equation}
\sigma'=\sigma-f^{\rm cl}_\pi
\label{shift}
\end{equation}
has been carried out. This shift leads to a mass term in the fermion
part of the Lagrangian
and to a constituent-quark mass in the chiral limit of
\begin{equation}
M=g f^{\rm cl}_\pi.
\label{constituent}
\end{equation}
Eq. (\ref{constituent}) is the same as Eq. (\ref{sigmamass}) except for the
difference that the NJL model provides a prediction, {\it viz.}
$g=\frac{2 \pi}{\sqrt{3}}$ for the meson-quark coupling constant whereas the
L$\sigma$M leaves this quantity undetermined. This is a beautiful example
for the fact that the NJL model has a higher predictive power than the
L$\sigma$M. Explicit symmetry
breaking is taken into account by the last term (\ref{ql1}) 
 which vanishes in the chiral
limit $m_\pi \to 0$.

 Eq.  (\ref{boson})  describes spontaneous chiral symmetry breaking
in terms of fields where a  Mexican hat potential 
\begin{equation}
V(\sigma,\pibf)=-\frac{\mu^2}{2}(\sigma^2+\pibf^2)+\frac{\lambda}{4}
(\sigma^2+\pibf^2)^2
\label{linearsigma}
\end{equation}
is introduced, parameterized by the mass
parameter 
$\mu^2>0$ and the self-coupling parameter $\lambda>0$. 
The minimum of the potential defines the vacuum expectation value of the 
$\sigma$ field $v^{\rm cl}_\sigma$ in the chiral limit via
\begin{equation}
v^{\rm cl}_\sigma\equiv f^{\rm cl}_\pi=\sqrt{\frac{\mu^2}{\lambda}}.
\end{equation}
There is no separate prediction 
for the two quantities
$\mu$ and $\lambda$ and, therefore, the mass of the $\sigma$ meson 
in the chiral
limit  $m^{\rm cl}_\sigma=\sqrt{2}\mu$
remains undetermined. This also is a well known fact for the electroweak 
Higgs boson, where instead of the $\sigma$ field we have to consider the Higgs
field and instead of the pions  three Goldstone bosons which are absorbed
into the longitudinal components of the electroweak gauge bosons.
Symmetry breaking described through a Mexican hat potential is customarily
named spontaneous symmetry breaking for obvious reasons. 
 The parameter $\lambda$ of the selfinteraction 
term is related to the coupling constants entering into the NJL model via
\begin{equation}
G=\lambda/(\sqrt{2}m^{\rm cl}_\sigma)^2, \quad g=\sqrt{\lambda/2}. \label{ql4}
\end{equation}
The r.h.s.  relation can be derived by expressing the $\sigma$ meson mass in the
chiral 
limit in the NJL model on the one hand and in L$\sigma$M on the other.
\begin{eqnarray}
&& \text{NJL}: \quad\, m^{\rm cl}_\sigma=2 g f^{\rm cl}_\pi,\\
&& \text{L$\sigma$M}: \quad m^{\rm
  cl}_\sigma=\sqrt{2}\mu=\sqrt{2\lambda}f^{\rm cl}_\pi. 
\label{tworelations}
\end{eqnarray}

The $SU(2)_L\times SU(2)_R$ symmetry of the L$\sigma$M is explicitly broken
if the potential $V(\sigma,\pibf)$ is made slightly asymmetric  by the addition
of the explicitly symmetry breaking  term given in (\ref{ql1}). This leads to
the vacuum expectation value $v_\sigma$ \cite{donoghue96}
\begin{equation}
v_\sigma=\sqrt{\frac{\mu^2}{\lambda}}+\frac{f^{\rm cl}_\pi
  m^2_\pi}{2\mu^2} 
\label{symbreak}
\end{equation} 
where $v_\sigma$ may be identified with the pion decay constant $f_\pi$.
The mass of the $\sigma$ meson may be written in the form \cite{alfaro73}
\begin{equation}
m^2_\sigma=2\lambda (f^{\rm cl}_\pi)^2+m^2_\pi.
\label{symbreak1}
\end{equation}
This equation (\ref{symbreak1}) does not predict the $\sigma$ meson mass as
long as 
the self-coupling parameter is undetermined. However, we 
have seen that the pion-quark coupling g is given by
\begin{equation}
g=\frac{2\pi}{\sqrt{3}} 
\label{symbreak2}
\end{equation}
leading to the self-coupling constant 
\begin{equation}
\lambda=2g^2=\frac{8\pi^2}{3}=26.3
\label{symbreak3}
\end{equation}
and to the mass parameter
\begin{equation}
\mu=\sqrt{\frac23}\, 2\pi f^{\rm cl}_\pi=460 \,\, \text{MeV}.
\label{parametermu}
\end{equation}

\section{Experimental confirmation of the predicted constituent-quark and 
$\sigma$-meson masses }

The magnetic moment, the electromagnetic polarizabilities  and the mass are
fundamental structure constants of the nucleon. On a quark level these
structure constants are closely related to each other. For the magnetic moment
and the mass this appears to be straightforward and well 
known (see e.g. \cite{ghalenovi14} and references therein). For the
electromagnetic polarizabilities this insight is new and has been explored 
in detail in recent papers (see \cite{schumacher13}). The main discovery is that the $\sigma$ meson 
as part of the constituent-quark structure has two important properties.
On the one hand the $\sigma$ meson mediates the interaction between the three 
quarks in the nucleon and the $\bar{q}q$ pairs in the QCD vacuum. On the
other hand the $\sigma$ meson has the property of coupling to two photons.
As a consequence of this the $\sigma$ meson participates in the Compton
scattering process and thus makes a dominant contribution to the electric
polarizability and takes care of the total diamagnetic polarizability. 
Additional contributions to the Compton scattering amplitudes and the
electromagnetic polarizabilities are provided by nonresonant and resonant
excitation processes of the nucleon. These latter contributions can be 
calculated
with high precision from photomeson CGLN amplitudes \cite{drechsel07}. 
As a consequence 
it becomes possible to precisely test the predictions of the contribution
the $\sigma$ meson to the Compton scattering amplitude and to the
polarizabilities. This also leads to a test of the predicted structure of the 
$\sigma$ meson.

\subsection{The magnetic moment of the nucleon}

It was one of the first evidences for the validity of the constituent-quark
model that it allows to correctly predict magnetic moments of the nucleon.
However, since there were
no predictions of the constituent-quark masses these quantities were
determined by adjusting the predicted magnetic moment for the proton to the
experimental value. Using these constituent-quark masses for a prediction
of the magnetic moment of the neutron  a good but not perfect agreement
with  the experimental value was obtained. The remaining discrepancy 
observed for the 
neutron led to the supposition that there also should be a component due to
meson exchange currents or configuration mixing in the nucleon
\cite{schumacher08}.

Our procedure is different from these previous approaches and was first 
described in a previous preprint  \cite{schumacher08}. Our  new approach 
makes use of the constituent-quark masses predicted by the 
NJL model on an absolute
scale. This procedure led to an excellent agreement simultaneously 
for the neutron and the
proton but also leaves room for some residual discrepancy of the order
of 1 -- 2 \%   which tentatively
may be attributed to meson exchange currents     or configuration mixing
\cite{schumacher08}. For sake of completeness 
we present an update of our previous considerations.

Symmetry considerations including the color degree of freedom lead to the
conclusion that the spins of identical  constituent quarks are parallel to
each 
other. Then the coupling of spin S=1 of the two identical constituent quarks to
S=1/2 of the nonidentical constituent quark leads to two contributions to the
total spin of the nucleon, as requested by the Clebsch-Gordan coefficients. 
Then the  magnetic moments of the nucleons are given by
\begin{eqnarray}
&& \mu_p=\frac43 \mu_u -\frac13\mu_d,\\
&& \mu_n=\frac43 \mu_d -  \frac13 \mu_u
\label{magneticmoment}
\end{eqnarray} 
in units of the nuclear magneton $\mu_N=e\hbar/2 m_p$. Constituent-quark
masses enter through the relations 
\begin{equation}
\mu_u=\frac23 \frac{m_p}{m_u},\quad \mu_d=-\frac13\frac{m_p}{m_d}.
\label{magnticmoment-2}
\end{equation}
On the basis of these equations we attempt a parameter-free prediction
of the magnetic moments of the nucleon. 
From the mass of the $\sigma$ meson $m_\sigma=666$ MeV (\ref{sigmamass-4})
we obtain
\begin{equation}
m_u=\frac12 m_\sigma - 2=331 \,\,\, \text{MeV},
\quad 
m_d=\frac12 m_\sigma + 2=335 \,\,\, \text{MeV}
\label{magneticmoment-3}
\end{equation}
where use is made of the fact that $m_d-m_u$ has been determined to be
approximately 4 MeV \cite{scadron13}.
This leads to the magnetic moments of the constituent quarks
\begin{equation}
\mu_u= 1.890,\quad \mu_d=-0.934
\label{quarkmagmom}
\end{equation}
and to predicted magnetic moments of the nucleon
\begin{equation}
\mu^{\rm theor}_p=2.831,\quad \mu^{\rm theor}_n=-1.875.
\label{nucleonmagmom}
\end{equation}
Comparing these values with the experimental magnetic moments of the nucleon
\begin{equation}
\mu^{\rm exp}_p=2.79285,\quad \mu^{\rm exp}_n= -1.91304
\label{magmoexo}
\end{equation}
we arrive at the very small differences $\Delta\mu=\mu^{\rm exp}-\mu^{\rm
  theor} $ of
\begin{equation}
\Delta\mu_p=-0.038,\quad \Delta\mu_n=-0.038.
\label{differences}
\end{equation}
These differences are the same for the proton and the neutron and amount to
$|(\Delta\mu/\mu)_p|=1.4\%$ and $|(\Delta\mu/\mu)_n|=2.0\%$, respectively.
The  interesting conclusion is  that the NJL model predicts constituent quark
masses which lead to  predictions  of magnetic moments of the nucleon
on a 1 -- 2\% level of agreement with the experimental values. We consider
this  as a first firm confirmation of the predicted constituent-quark masses.
A second confirmation is provided by Compton scattering and the  
polarizabilities of the nucleon as described in the next subsection.

The  constituent quark masses $m_u=331$ MeV and $m_d=335$ MeV are the
same as  the mass parameters used 
in other work \cite{ghalenovi14}, i.e. $m_u=330$ MeV and $m_d=335$ MeV.
The difference between our present work and \cite{ghalenovi14} is that our
numbers  are derived on an absolute scale by applying the NJL model.

\subsection{Compton scattering and polarizabilities of the nucleon}

A nucleon in an electric field ${\bf E}$ and  magnetic field ${\bf H}$ obtains
an induced electric dipole moment ${\bf d}$ and magnetic dipole 
moment ${\bf m}$
given by 
\begin{eqnarray}
&&{\bf d}= 4\pi \,\alpha \,{\bf E}, \label{electric}\\
&&{\bf m}= 4\pi\, \beta \,{\bf H}, \label{magnetic}
\end{eqnarray}
in a unit system where the electric charge $e$ is given by  
$e^2/4\pi =\alpha_{em}=1/137.04$. Eqs. (\ref{electric})
and (\ref{magnetic}) may be understood as the response of the nucleon 
to the fields provided by a real or virtual photon and it is evident
that we need a second photon to measure the polarizabilities. This may be
expressed through the relation
\begin{equation}
H^{(2)}=-\frac12\,4\pi\,\alpha\,{\bf E}^2 -\frac12\,4\pi\,\beta\,{\bf H}^2
\label{energy}
\end{equation}
where $H^{(2)}$ is the energy change in the electromagnetic field due to the
presence of the nucleon in the field. 

For particles with a spin  spinpolarizabilities have to be taken into
account. These quantities differ from the polarizabilities by the fact that
the two photons measuring these quantities are in perpendicular planes of
linear polarization. In this case $\gamma_0$ denotes the spinpolarizability
for Compton scattering in the forward direction and $\gamma_\pi$ the
spinpolarizability for Compton scattering in the backward direction.

\subsubsection{Diagrammatic representation of Compton scattering}

\begin{figure}[h]
\begin{center}
\includegraphics[width=0.5\linewidth]{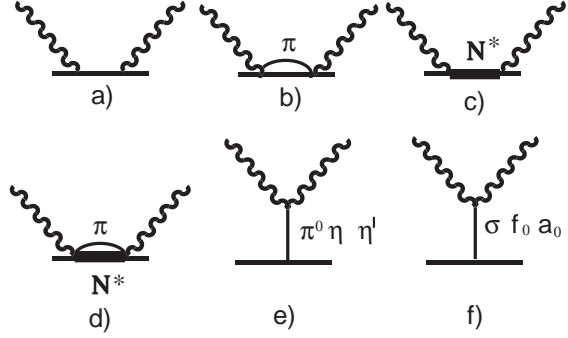}
\end{center}
\caption{Graphs  of nucleon Compton scattering. a) Born term,
b) single-pion photoproduction, c) resonant excitation of the nucleon
and d) two-pion photoproduction. 
The crossed graphs 
accompanying the graphs a) to d) are not shown. The $t$-channel graphs in e)
and f) correspond to pointlike singularities due to the pseudoscalar meson
$\pi^0$, $\eta$ and $\eta'$ and the scalar mesons $\sigma(600)$, $f_0(980)$
and $a_0(980)$. These  graphs are shown for illustration whereas
quantitative predictions are based on dispersion theory.}
\label{comp-graph}
\end{figure}
The method of predicting  polarizabilities is based on dispersion theory
applied to partial photoabsorption cross-sections. 
Nevertheless, it is of interest
for illustration to show the Compton scattering process in terms of
diagrams as carried out in Figure \ref{comp-graph}.
We see that in addition to the nonresonant single-pion channel (b)
and the resonant channel (c) we have three further contributions.
Graph (d) corresponds to Compton scattering via the transition
$\gamma N\to \Delta \pi$ to the intermediate state.   
This represents  the main component of those photomeson processes 
where two pions are in the final state. Graph e) is the well-known
pseudoscalar $t$-channel graph where in addition to the dominant
$\pi^0$ pole contribution also the small contributions due to $\eta$
and $\eta'$ mesons are taken into account. Graph f) depicts the
scalar counterpart of the pseudoscalar pole term e). At  first sight
these scalar pole-terms appear inappropriate because the scalar particles
and especially the $\sigma$ meson have a large width when observed on shell.  
However, this consideration overlooks that in a Compton scattering process
the scalar particles are intermediate states of the scattering process
where the broadening due to the $\pi\pi$ final state is not effective. 
This leads to the conclusion that the scalar pole terms indeed provide the
correct description of the scattering process.

\subsubsection{Predicted polarizabilities obtained for the $s$-channel}

Using  sum rules (see \cite{schumacher13}) and applying them to 
 photoabsorption cross sections the
polarizabilities  and spin polarizabilities given in Table 
\ref{polresults1} are obtained.  Lines 2 contain the contributions of the
$P_{33}(1232)$ nucleon resonance. This resonance is exited from the ground
state via a spin-flip transition and thus is the main source of paramagnetism
of the nucleon.
The nonresonant electric-dipole  contribution due to the  $E_{0+}$ 
CGLN amplitude \cite{drechsel07} is  given
in line 3. For the electric polarizability this corresponds to the 
main  part of the ``pion-cloud''
contribution. 
 Line 4 contains the total $s$-channel contribution.  

 \begin{table}[h]
\caption{Electric ($\alpha$) and magnetic $(\beta)$ polarizabilities 
and spin polarizabilities
 for the forward $(\gamma_0)$ and backward $(\gamma_\pi)$ directions.
Line 2: Resonant contributions due to the $P_{33}(1232)$ resonance.
Line 3 : Single-pion nonresonant contribution due to the $E_{0+}$ CGLN
amplitude \cite{drechsel07}.  Line 4: Total
$s$-channel components of the polarizabilities.
The electric and magnetic
  polarizabilities are in units of $10^{-4}$fm$^3$, the 
spin polarizabilities  in units of $10^{-4}$fm$^4$.}
\begin{center}
\begin{tabular}{l|ll|ll|ll|ll}
\hline
&$\alpha_p$&$\beta_p$&$\alpha_n$&$\beta_n$&
$\gamma^{(p)}_0$&$\gamma^{(n)}_0$&$\gamma^{(p)}_\pi$&$\gamma^{(n)}_\pi$\\ 
\hline
$P_{33}(1232)$&$-1.07$&$+8.32$&$-1.07$&$+8.32$ &
$-3.03$&$-3.03$&$+5.11$&$+5.11$\\
$E_{0+}$&$+3.19$&$-0.34$&$+4.07$&$-0.43$&
$+2.47$&$+3.18$&$+3.75$&$+4.81$\\
$s$-channel&$+4.48 $&$+9.44 $&$+5.12 $&$+10.07 $&
$-0.58 $&$+0.38 $&$+8.47  $&$+10.00 $\\
\hline
\hline
\end{tabular}
\label{polresults1}
\end{center}
\end{table}

\subsubsection{Contributions to the polarizabilities due to the mesonic
  structure of the constituent quarks}

Contributions to the polarizabilities stemming from the mesonic structure
of the constituent quarks enter into dispersion theory via the $t$-channel.
The main parts of  these $t$-channel contributions are due to the $\pi^0$
meson in case of the spinpolarizability $\gamma_\pi$ 
and due to the $\sigma$ meson
in case of the electromagnetic polarizabilities $\alpha$ and $\beta$. For the
present article the results obtained for the $\sigma$ meson  are of special
interest because they contain a direct proof that  the 
$\sigma$ meson as part of the constituent-quark structure is  well understood.

 The main part of the $t$-channel component
of the backward spinpolarizability is given by
\begin{eqnarray}
&&|\pi^0\rangle=\frac{1}{\sqrt{2}}
(-|u\bar{u}\rangle+|d\bar{d}\rangle),\quad
{\cal M}(\pi^0\to\gamma\gamma)=\frac{\alpha_{em}N_c}{\pi f_\pi}
\left[-\left(\frac23\right)^2
+\left(\frac{-1}{3}\right)^2\right], \label{sp1}\\
&&{\gamma_\pi^t}_{(p,n)}=\frac{g_{\pi^0 NN} \quad
{\cal M}(\pi^0\to\gamma\gamma)}
{2\pi m^2_{\pi^0} m}\tau_3= -46.7\,\tau_3\,\,\, {10}^{-4}{\rm fm}^4.
\label{sp2}
\end{eqnarray}
Analogously, we obtain for the main $t$-channel parts of the electric
($\alpha$) and magnetic ($\beta$) polarizabilities the relations 
\begin{eqnarray}
&&|\sigma\rangle=\frac{1}{\sqrt{2}}(|u\bar{u}\rangle
+|d\bar{d}\rangle), \quad
{\cal M}(\sigma\to\gamma\gamma)=\frac{\alpha_{em}N_c}{\pi f_\pi}
\left[\left(\frac23\right)^2
+\left(\frac{-1}{3}\right)^2\right],\label{sp3}\\
&&(\alpha-\beta)^t_{p,n}=\frac{g_{\sigma NN}{\cal  M}(\sigma\to\gamma\gamma)}
{2\pi m^2_\sigma}= 15.2\,\,\, {10}^{-4}{\rm fm}^3,\quad
(\alpha+\beta)^t_{p,n}=0,\\
&&\alpha^t_{p,n}=+7.6\,\,\, {10}^{-4}{\rm fm}^3,\,
\beta^t_{p,n}=-7.6\,\, {10}^{-4}{\rm fm}^3\label{sp4},
\end{eqnarray}
where use is made of $g_{\pi NN}=g_{\sigma NN}=13.169\pm 0.057$ 
\cite{bugg04}
and
$m_\sigma=666$ MeV as predicted by the NJL model. The sign convention
used in the $q\bar{q}$ structure of the $\pi^0$ meson follows from 
\cite{close79}. It has the advantage of correctly predicting the sign of the
$\pi^0$-pole contribution.
These main contributions to the polarizabilities $\alpha$, $\beta$ and
$\gamma_\pi$ have to be supplemented by the $s$-channel components 
and by the small components due to the scalar mesons $f_0(980)$ and 
$a_0(980)$ in case of the polarizabilities $\alpha$ and $\beta$ and 
due to the pseudoscalar meson $\eta$ and $\eta'$ in case of $\gamma_\pi$.

The numerical evaluation
of these  contributions has been described in detail in previous papers
\cite{schumacher07b,schumacher09}. Here we 
give a summary of the final predictions and the experimental values to compare
with in Table \ref{tab3} and \ref{tab4}. 
\begin{table}[h]
\caption{Backward spin polarizabilities for the proton and the neutron from 
\cite{schumacher11a} and Table \ref{polresults1}} 
\begin{center}
\begin{tabular}{l|ll}
spin polarizabilities & $\gamma^{(p)}_\pi$ & $\gamma^{(n)}_\pi$\\
\hline
$\pi^0$ pole & -46.7 & +46.7\\
$\eta$ pole & +1.2 & +1.2\\
$\eta'$ pole & +0.4 & +0.4\\
\hline
const. quark structure&$-$45.1 &+48.3\\
nucleon structure & +8.5 & +10.0 \\
\hline
total predicted& $-$36.6 & +58.3\\
exp. result & 
$-$($36.4\pm 1.5$)& 
+($58.6\pm 4.0$)\\
\hline
&unit 10$^{-4}$ fm$^4$
\end{tabular} \label{tab3}
\end{center}
\end{table}
\begin{table}[h]
\caption{Polarizabilities for the proton and the neutron from 
\cite{schumacher09}
and Table \ref{polresults1}. The total predicted values given in parentheses
are obtained in case the $f_0(980)$ and $a_0(980)$ contributions are 
disregarded.} 
\begin{center}
\begin{tabular}{l|ll|ll}
 & $\alpha_p$ & $\beta_p$ & $\alpha_n$ & $\beta_n$\\
\hline
$\sigma$ pole & +7.6 & $-7.6$ & +7.6 &$-7.6$\\
$f_0$ pole & +0.3 & $-0.3$ & +0.3 & $-0.3$\\
$a_0$ pole & $-0.4$ & +0.4& +0.4 & $-0.4$\\
\hline
const. quark structure& +7.5 & $-7.5$
& +8.3 & $-8.3$\\
nucleon structure  & +4.5 & +9.4 & +5.1 & +10.1\\
\hline
total predicted& +12.0 (+12.1) & +1.9 (+1.8) & +13.4 (+12.7) & +1.8 (+2.5)\\
exp. result  & 
+($12.0\pm 0.6$)
& +($1.9\mp 0.6$)& 
+($12.5\pm 1.7$)&
+($2.7\mp 1.8$) \\
\hline
&unit 10$^{-4}$ fm$^3$
\end{tabular}\label{tab4}
\end{center}
\end{table}

In Table \ref{tab3} we find an excellent agreement between the experimental
and predicted backward spin polarizabilities. This proves that both, the
experimental and theoretical procedures are correct.

The most interesting feature  of the polarizabilities  given in Table
\ref{tab4} is the strong cancellation of the paramagnetic polarizabilities 
which is mainly due to the $P_{33}(1232)$ resonance by the diamagnetic
term which is solely due to the constituent-quark structure. This
constituent-quark structure component is mainly due to the $\sigma$-meson
pole contribution with minor corrections due to the $f_0(980)$ and
$a_0(980)$ scalar mesons. The predictions obtained for these latter
contributions take into account the information we have from the two-photon
width but remain uncertain to some extent. Therefore, it may be appropriate 
not to include the respective predictions into the final result. The effects
of these two options  may be seen in line 7 of Table \ref{tab4}
where the numbers in parentheses do not include the contributions from
the  $f_0(980)$ and $a_0(980)$ scalar mesons.  Apparently the effects of
the $f_0(980)$ and $a_0(980)$ scalar mesons are small and do not lead to a
major uncertainty in the predictions as stated before.

A further interesting conclusion concerns the  two-photon width of the
$\sigma$ mesons which can be calculated from the data given in 
(\ref{sp3}). Given the $\sigma$ meson mass of 666
MeV the amplitude for the  $\sigma\to\gamma\gamma$ decay for $N_c=3$ and
$f_\pi=92.42\pm 0.26$ MeV is
\begin{equation}
{\cal M}(\sigma\to\gamma\gamma)=\frac{5\,\alpha_{\rm em}}{3\pi f_\pi}=0.042
\,\,\text{GeV}^{-1}
\label{num1}
\end{equation}
and the decay width
\begin{equation}
\Gamma_{\gamma\gamma}=\frac{m ^3_\sigma}{64\pi}
|{\cal M}(\sigma\to\gamma\gamma)|^2
=2.6\,\, \text{keV}.
\label{num2}
\end{equation}

From Table \ref{tab4} we obtain that the
predicted and the experimental electric polarizabilities $\alpha_p$ 
of the proton are in excellent agreement with each other. Furthermore, the
experimental quantity $\alpha_p$ has an experimental error of $\pm 5\%$
which may be used
to calculate the experimental error of the two-photon width  
$\Gamma_{\gamma\gamma}$ of the $\sigma$ mesons. This rather straightforward
consideration leads to $\Gamma_{\gamma\gamma}= 2.6\pm 0.3$ keV
\cite{schumacher10}. This result obtained from the quark model and from
the electric polarizability $\alpha_p$ of the proton is in agreement 
with the two-photon width extracted from data on the 
$\gamma\gamma\to \pi\pi$ reaction \cite{schumacher10},
leading to a further proof that the present 
quark model description of the structure of the $\sigma$ meson is correct.

\subsubsection{Interference between nucleon and constituent-quark structure
  Compton scattering}

The most direct observation of the $\sigma$ meson as part of the
constituent-quark structure has been achieved via Compton scattering by the
nucleon carried out in the photon energy-range between 400 and 700 MeV
at the electron accelerator MAMI (Mainz) \cite{galler01,wolf01}. In
this  energy range the differential cross section for Compton scattering by
the  nucleon is dominated by an interference effect between  a contribution
from the $P_{33}(1232)$ resonance and the $\sigma$ meson pole contribution.
The predicted $\sigma$ meson pole contribution depends on the $\sigma$ meson
mass entering into the calculation. This means that the mass of the $\sigma$
meson can be determined by adjusting the predicted Compton differential 
cross-section to the  experimental one. 

This will be carried out in the following
by investigating the energy dependence of the polarizabilities
$(\alpha-\beta)$.  
\begin{figure}[h]
\includegraphics[width=0.4\linewidth]{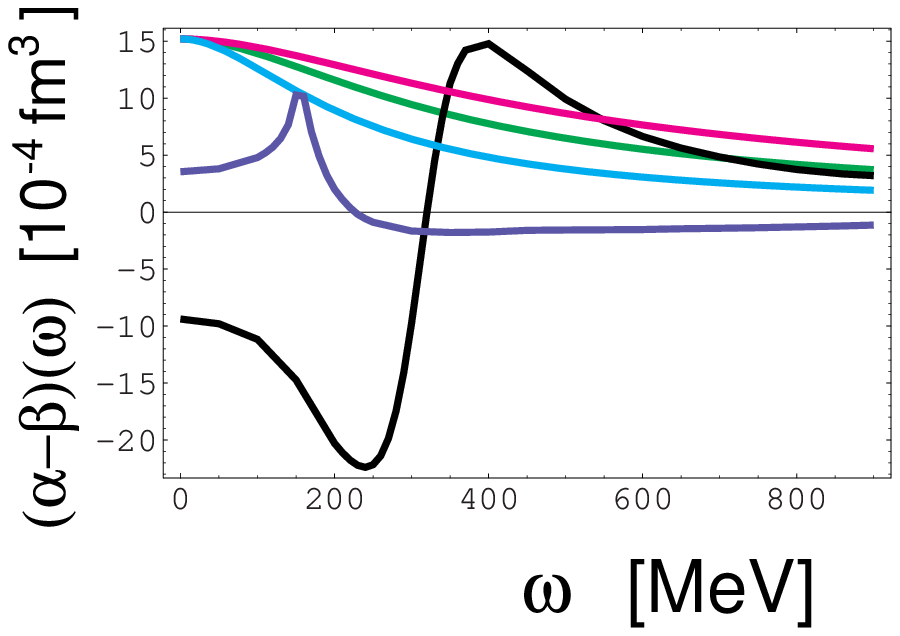}
\includegraphics[width=0.5\linewidth]{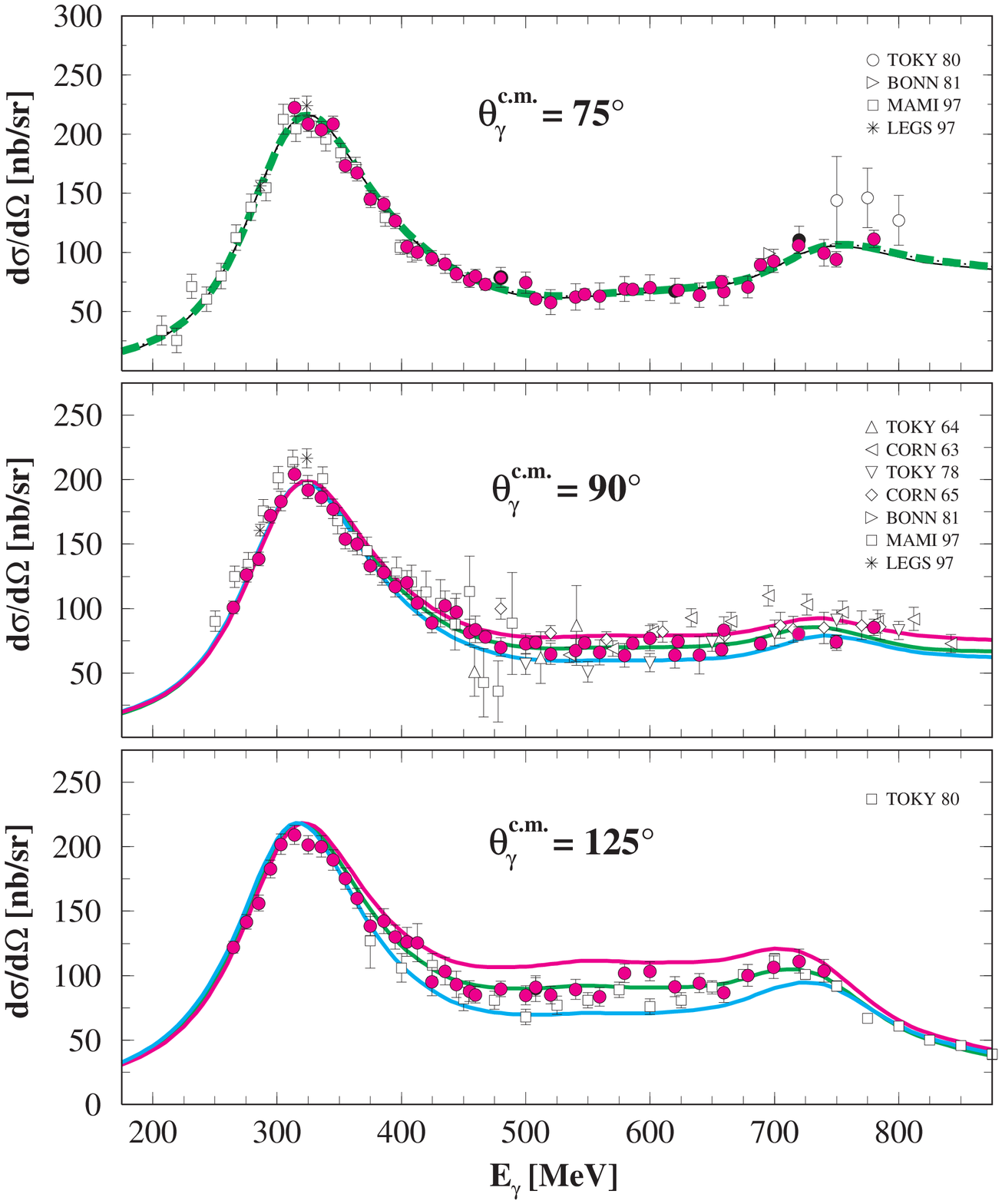}
\caption{Left panel: Difference of generalized polarizabilities 
$(\alpha-\beta)(\omega)$
versus photon energy $\omega$. Solid curve starting at
$(\alpha-\beta)(0)=-9.4$:
contribution of the $P_{33}(1232)$ resonance. Solid curve starting at
$(\alpha-\beta)(0)=+3.5$: Contribution of the nonresonant $E_{0+}$ amplitude.
Curves starting at $(\alpha-\beta)(0)=+15.2$: $t$-channel contribution of the
$\sigma$ meson calculated for different $\sigma$-meson masses. Upper curve
(dark grey or red): $m_\sigma=800$ MeV. Middle curve (grey or green):
$m_\sigma=600$ MeV. Lower curve (light grey or blue): $m_\sigma= 400$ MeV.
Not shown is the contribution of the $D_{13}(1520)$ resonance which cancels
the $E_{0+}$ contribution in the relevant energy range from 400 to 700 MeV.
Right panel: Differential cross sections for Compton scattering by the proton
versus photon energy. The three panels contain data corresponding to the
cm-angles of $75^\circ$, $90^\circ$ and $125^\circ$. The three curves 
are calculated for different mass parameters $m_\sigma=800$ MeV (upper, dark
grey or red), 600 MeV (center, grey or green) and 400 MeV (lower, light grey
or blue).\label{f1}
}
\end{figure}
The results obtained  are shown in the
left panel of Figure \ref{f1}.  It is clearly demonstrated that there is a
constructive interference in the energy range from 400 to 700 MeV
between the contribution from the $P_{33}(1232)$
resonance and from the $t$-channel due to the $\sigma$ meson as part of the
constituent-quark structure. The sum of the two contributions depends on the
mass  
of the $\sigma$ meson adopted  in the calculation and, therefore, may be used
for a determination of the $\sigma$-meson mass $m_\sigma$ from the differential
cross section for Compton scattering. This is outlined in the right panel of
Figure \ref{f1} where the effects of the $\sigma$-meson mass are most
prominent in the backward direction, as expected. The best agreement is
obtained for $m_\sigma=600\pm 70$ MeV, being compatible with our present knowledge
of the ``bare'' mass of the $\sigma$ meson. As pointed out before
 \cite{schumacher10} the experimental data shown in Figure \ref{f1} may be
 interpreted as a direct observation of the $\sigma$ meson while being part
 of the constituent-quark structure. Furthermore, this experiment provides us
 with the 
first and only direct determination of the ``bare'' mass of 
the $\sigma$ meson,
whereas in all the other observations of the $\sigma$ meson 
the Breit-Wigner parameters or
the position of the
pole on the second
Riemann sheet are determined.

\section{The masses of nucleons and nuclei}

When we speak of the mass of visible matter we have to consider the masses of
nuclei and how they are composed of nucleons, and the masses of nucleons
and how they are composed of constituent quarks.
 Nuclei and nucleons have in common
that they contain particles which are held together by interparticle 
forces. In case of nuclei these internucleon forces are of  mesonic 
origin where the largest attractive part is provided by the $\sigma$ meson. 
In addition there are minor effects due to Coulomb forces 
acting between protons.
In case of nucleons we have two components one mesonic component analogous to
the nuclear case \cite{glozman96} and one gluonic component  provided by 
the color charges  of the quarks. 
Furthermore, minor effects due to Coulomb forces are expected.
The color charges are responsible for one essential
difference between nuclei and nucleons. In case of nuclei the nucleons
can be extracted from the nucleus if the energy transfer is high enough.
This is different in case of nucleons where the color charges lead
to confinement. Single quarks cannot be extracted from the nucleon
but only quark-antiquark pairs bound together to form a meson. At low
energies these mesons are pions. 

Except for a binding energy  the mass of the nucleus is given by the 
masses  of the nucleons. In extended  nuclei the binding energies amount to 8
MeV 
per nucleon on the average.
For the light
nuclei $^3_1$H and $^3_2$He the corresponding numbers are 2.83 MeV and 2.57
MeV, respectively. These numbers are well explained in terms of mass formulae
developed in nuclear physics.
In order to understand the
missing 98\% of visible matter we, therefore, have to explain the masses
of the nucleons. In the foregoing subsections we have shown that the
constituent-quark masses $m_u=331$ MeV and $m_d=335$ MeV are well founded
values. Without mass reduction due to the
binding energy  the mass of the proton would be $m^0_p=997$ MeV
and the mass of the neutron $m^0_n=1001$ MeV. Comparing these numbers with
the experimental nucleon masses leads to the binding energies per constituent
quark given in Table \ref{quarkbinding}.
\begin{table}[h]
\caption{Binding energy B per constituent quark (A=3) in the two isospin 
partners
p and n.}
\begin{center}
\begin{tabular}{c|c|c}
\hline
nucleon& p& n\\
\hline
B/A & 19.6 MeV& 20.5 MeV\\
\hline
\end{tabular}
\end{center}
\label{quarkbinding}
\end{table}
The binding energies  obtained for the quarks in the nucleon are about
a factor 7.5 larger than the corresponding numbers for the nucleons
in the light nuclei $^3_1$H and $^3_3$He. This result is very plausible
because of the smaller distances in the nucleon and because of gluonic
components in the inter-quark forces.
In a tentative attempt to understand the difference of the numbers 
obtained for B/A in the proton and the neutron we write down the interquark
Coulomb energy  in the form
\begin{equation}
U=\sum_{i,j,i< j}\frac{e_i e_j}{r_{ij}}\alpha_{\rm em} \hbar c.
\label{Coulomb}
\end{equation}
Then, with $\langle r_{ij} \rangle \approx 0.3$ fm we arrive at
$U_p\approx 0$ MeV and $U_n\approx -1.6$ MeV. This tentative consideration
explains the difference of  the B/A values for the proton and the neutron 
as being partly due to a Coulomb attraction in the neutron.  
Though there is no complete interpretation of the B/A values given in
Table \ref{quarkbinding}, the smallness of these residual mass contributions 
demonstrates that on the whole the masses of the nucleons are quantitatively
predicted by the JNL model on a high level of precision.

\section{Discussion and conclusions}

In the foregoing we have shown that chiral symmetry breaking as provided
the NJL model simultaneously explains the fundamental structure constants
of the nucleon, {\it viz.} the magnetic moment, the electromagnetic
polarizabilities and the mass. The magnetic moment and the electromagnetic
polarizabilities are used as a proof that the predictions obtained for the
constituent-quark masses are correct. These constituent-quark masses, when
added together reproduce the experimental nucleon mass except for a binding
energy of  $\sim 20$ MeV per constituent quark. This very satisfactory result
proves that the concept of mass generation as provided  
by the NJL model is correct. 

The discovery of the electroweak Higgs boson has led to models of the vacuum
where an overall Higgs field is assumed to exist. For a mass of the
Higgs boson of 126 GeV this Higgs field is expected to be elementary, i.e. 
not composed of fermion-antifermion pairs. The reason for this conclusion
is  that Higgs bosons composed of $t\bar{t}$ pairs or
techniquark-antitechniquark pairs are predicted to have higher masses.
In parallel to this, for the strong-interaction counterpart  an 
overall $\sigma$ field may be assumed to exist in the QCD vacuum which
 is composed $u\bar{u}$ and $d\bar{d}$ pairs forming  the structure
$\sigma= (u\bar{u} + d\bar{d})/\sqrt{2}$. Then the generation
of the mass of the constituent quarks may be understood in terms a $q\bar{q}$
condensate attached to the current quarks. In this condensate the 
$q\bar{q}$ pairs are ordered to form mesons, like the $\pi$ meson isospin
triplet and the $\sigma$ meson.

\end{document}